\documentclass[
reprint,
superscriptaddress,
amsmath,amssymb,
prb,
]{revtex4-2}

\usepackage{dcolumn}
\usepackage{bm}
\usepackage{graphicx}
\usepackage{color}
\usepackage{amsmath,bm}
\usepackage{siunitx}

\begin{document}

\title{Electronic effects in radiation-induced collision cascades in nickel}

\author{Andrea E. Sand}
\email{andrea.sand@aalto.fi}
\affiliation{Department of Applied Physics, Aalto University, 00076 Aalto, Espoo, Finland}
\author{Glen P. Kiely}
\affiliation{Department of Applied Physics, Aalto University, 00076 Aalto, Espoo, Finland}
\author{Artur Tamm}
\affiliation{Institute of Physics, University of Tartu, 50411 Tartu, Estonia}
\affiliation{Quantum Simulations Group, Physics Division, Lawrence Livermore National Laboratory, Livermore, California 94551, USA}
\author{Alfredo A. Correa}
\affiliation{Quantum Simulations Group, Physics Division, Lawrence Livermore National Laboratory, Livermore, California 94551, USA}


\date{\today}

\begin{abstract}
The accurate treatment of electronic effects in multi-million atom simulations of radiation-induced collision cascades is crucial for reliable predictions of primary radiation damage. 
In this work, we explore the performance of a recently developed two-temperature molecular dynamics model implementing an electron density-dependent coupling of electronic and atomic subsystems for cascade simulations in nickel.
We show that the parameter-free model realistically captures the instantaneous energy losses during all stages of the highly non-equilibrium cascade process.
Simulations predict two distinct coupling regimes, corresponding to the rapid electronic stopping energy losses in the early stages of the cascade and to the electron-phonon coupling mechanism in the later stages, without the use of separate coupling terms.
The intermediate stage of the cascade dynamics displays a complex energy transfer between the subsystems, which cannot be validated by comparison to either electronic stopping or electron-phonon coupling theories. 
We therefore compare the predicted atomic mixing, which is sensitive to the energy losses during the intermediate cascade stage, with experimental ion beam mixing measurements. 
We find good agreement with the experiments, validating the coupling model for the intermediate stage of cascades. 
Predictions of final defect numbers and cluster sizes are found in line with the predictions of conventional electronic stopping-based methods, while significantly reducing the theoretical uncertainty in the predictions of conventional models stemming from arbitrary choices of thresholds for different coupling terms. 
Our results represent a notable improvement in cascade damage predictions in nickel, providing validation of the electron density-dependent coupling model for radiation damage simulations in general.
The results lead us to propose an interpretation of the dissipation in an intermediate regime of velocities where we find a non-linear dissipation.
\end{abstract}

\maketitle

\section{\label{sec:intro}Introduction}

The interaction of energetic ions and neutrons with materials is relevant for many technological applications, ranging from material modification by ion beams to nuclear energy production.
Displacement damage, caused by energetic particles colliding with lattice atoms, gives rise to changes in the mechanical and physical properties of irradiated materials over time.
Quantitative predictions of the displacement damage, or primary radiation damage, constitute the first step in predicting the response of materials to irradiation.

Primary radiation damage forms on picosecond timescales, making the process difficult to observe experimentally.
Hence, modelling is paramount for understanding the initial defect formation, with molecular dynamics (MD) providing an ideal tool for simulating the rapid atomic-scale process.
However, MD relies on the Born-Oppenheimer approximation and the assumption that electrons are always in their stationary state~\cite{Allen}.
Nevertheless, it is well known that energetic ions in matter lose energy through inelastic collisions with electrons, leading to electronic stopping.
In addition, the possible impact of electron-phonon coupling on the cooling rate of the thermal spike phase of cascades has been debated for decades \cite{Fly88,Car89,Fin91,Nor98}.
Furthermore, MD does not account for other direct effects of the electronic system, such as the electron-dominated thermal conduction in metals. 
Accordingly, extensions to MD have been developed to better capture the full effects of radiation damage dynamics. 
Traditionally, the explicit treatment of electronic effects in cascade simulations has consisted of implementing the two distinct and well-defined processes of electronic stopping and electron-phonon coupling.

Electronic stopping has been incorporated in MD simulations of collision cascades mainly through a retarding force imposed on energetic atoms.
In general, the magnitude of the retarding force experienced by mid- to low-energy ions is expected to be proportional to the velocity of the ion \cite{Fer47,Lin63,Ech81}, and the magnitude of the retarding force is typically set to agree with the semi-empirical predictions of SRIM \cite{SRIM}.
However, both experimental measurements~\cite{Val03,Mar09,Mar09b} and theoretical models~\cite{Pru07,Lim16,Qua16} indicate that band structure and non-linear effects become increasingly important in the low energy limit.
Direct experimental measurements in the low-energy regime are challenging, and data is scarce; thus, uncertainties exist in the treatment of energy losses for cascade atoms with low recoil energy.  
In addition, to avoid unphysical quenching of thermal modes in MD simulations, it is necessary to set a lower velocity cut-off on the retarding force. 
Although the choice of the cut-off value shows a quantitative effect on predictions~\cite{San13,San15}, there is no clear consensus regarding what this value should be. 
Tight-binding calculations~\cite{Pag09} indicate an optimal threshold at \(1~\mathrm{eV}\), which is supported by MD results in Fe~\cite{Bjo09d}, while comparisons of MD simulations with TEM observations in W~\cite{San13, San18b} and with ion beam mixing (IBM) experiments~\cite{Kim88} in Ni, Pd and Pt~\cite{San15} suggest that a threshold at \(10~\mathrm{eV}\) gives more accurate predictions. 
Another approach has been to set the cut-off at twice the material-specific cohesive energy~\cite{Zar16}.

Closer to equilibrium, ions and electrons couple through the electron-phonon (e-ph) interaction. 
Early MD studies implemented e-ph coupling in cascade simulations purely as a damping force~\cite{Fin91,Gao98}, an approach that also required the use of a velocity cut-off to avoid unphysical quenching. 
In more recent models \cite{Duf07,Duf09,Zar14,Zha13}, a velocity cut-off on the e-ph coupling has been avoided by the use of a two-temperature molecular dynamics (TTMD) framework, where energy feedback from the electronic system is implemented as a stochastic term that couples an electronic (continuum) heat bath to the atomic system.
The model is based on the concept of Langevin dynamics, where the stochastic force is rationalized near equilibrium and interpreted as originating from thermal fluctuations in the electronic degrees of freedom.
The model, initially developed for irradiation damage simulations by Duffy and Rutherford~\cite{Duf07, Rut07}, employs two distinct coupling parameters for the electronic stopping and e-ph coupling. A velocity threshold is applied to the electronic stopping as in the purely friction model, and a threshold in time has been suggested for the e-ph coupling~\cite{Zar17b} motivated by the longer time scale of the e-ph coupling compared to early cascade dynamics.
Recent parametric studies~\cite{Jar20,Zho20,Cui20} clearly show that these thresholds and the choice of cut-off values affects predictions of cascade simulations in the TTMD model as well, yet a definitive demonstration of physically sound values is lacking.

Caro and Victoria~\cite{Car89} conjectured early on that the same physics must underlie the phenomena of electronic stopping and electron-phonon coupling, with the difference being only a matter of degree.
They argued that the electronic density locally around atoms should regulate the strength of the coupling and could be used as a proxy, allowing modelling of both processes within a unified framework. 
With the advent of time-dependent density functional theory (TDDFT) methods, it has become possible to investigate the local electron density dependence of energy losses from first principles \cite{Cor12,Car15,Cor18}, and with this, parametrize the proposed model.
The approach results in a threshold-free dynamic coupling parameter that is a direct function of the local environment of each ion.
In all cases, the instantaneous coupling is quantitatively interpreted as the proportionality factor between ion dissipation forces and their velocities.
Using this method, a coupling function has been obtained for a number of metals \cite{Car19}.
The electron-density dependent coupling scheme in a two-temperature framework has also been recently implemented as a user plug-in (USER-EPH) for the LAMMPS code \cite{Tam18,Tam19}.
Within this approach, the effective electronic stopping power is determined from first principles, based on specific real-time (rt)TDDFT calculations, which have shown good agreement with experiments~\cite{San19,Gu20}.

Here, we demonstrate the applicability of this model, which we will refer to as the EPH model, for full cascade simulations where both the electronic stopping and e-ph coupling regimes are active.
We approach the question through a study of nickel, as Ni-based alloys are used in a number of demanding nuclear applications \cite{Gri20} due to their corrosion resistance and good mechanical properties at high temperatures.
Furthermore, many computational studies of radiation damage in Ni have been carried out using a wide variety of methods, including the TTMD model of Duffy and Rutherford~\cite{Zar17b}, as well as the purely friction model~\cite{Nor98,San15}.
We compare the cascade evolution, the predicted ion beam mixing, and the number of defects and defect clusters in cascade simulations employing different dissipation models. 
Results from the EPH model are compared to the model employing only an electronic stopping friction term in conventional MD, with no explicit electronic subsystem, and with the electronic stopping power (ESP) determined by SRIM~\cite{SRIM}. 
For the friction-based method, two different cut-off velocities corresponding to the kinetic energies of \(1\) and \(10~\mathrm{eV}\) are applied and the theoretical uncertainties discussed. 
In the following, we denote these models by the labels `EPH', `ESP-1eV' and `ESP-10eV', respectively.

\section{Methods}

\subsection{Dissipation models}

The EPH model is implemented as a generalized two-temperature molecular dynamics formalism and is described in detail in Refs.~\cite{Tam18,Car19,Tam19}.
Briefly, the model implements the coupled dynamics of electronic and atomic subsystems.
The atom dynamics are governed by equations of motion of the form
\begin{equation}
m_i\mathbf{a}_i = \mathbf{F}_i - \bm{\sigma}_i + \bm{\eta}_i
\end{equation}
where 
\(\mathbf{F}_i\) is the force on atom \(i\) from neighboring atoms determined by the interatomic potential (\(-\nabla U(\{\mathbf{r}_i\}\)) of the classical MD, 
\(\bm{\sigma}_i\) is a friction term, 
and \(\bm{\eta}_i\) is a stochastic force determined by a generalized form of Langevin dynamics \cite{Tam18}.
Stochastic forces naturally have local statistical correlations that follow corresponding tensorial relations.
The (tensorial) friction term is proportional to the relative velocities of the ions in the local neighborhood and is defined by 
\begin{equation}
\bm{\sigma}_i = \sum_j \mathbf{B}_{ij} \mathbf{v}_j
\end{equation}
where each \(3\times 3\) tensor \( \mathbf{B}_{ij} = \sum_k \mathbf{W}_{ik}\mathbf{W}_{jk}^\mathrm{T}\) is positive-definite, and \(\mathbf{W}_{ij}\) is given by
\begin{equation}
\mathbf{W}_{ij} = \begin{cases}    -\alpha_j \frac{\rho_i(r_{ij})}{\bar\rho_j} \mathbf{e}_{ij}\mathbf{e}_{ij}, & (i\neq j)\\
    \alpha_i \sum_{k\neq i} \frac{\rho_k(r_{ik})}{\bar\rho_i} \mathbf{e}_{ik}\mathbf{e}_{ik}, & (i = j)
\end{cases}
\end{equation}
where \(\alpha_i = \alpha(\bar\rho_i)\), and \(\bar\rho_i\) is the electronic density at the position of atom \(i\), given by the sum of contributions of spherical densities from neighboring atoms.
Thus, the overall magnitude of the friction term is set by \(\alpha^2(\rho)\), which is parametrized based on rt-TDDFT calculations.
It has dimensions of \(\mathrm{[F][T]/[L]}\), or \(\mathrm{[M]/[T]}\), while the other factors are dimensionless.
The EPH-model is implemented as an open-source user package (USER-EPH) \cite{EPHLAMMPS} for LAMMPS \cite{LAMMPS}.

The electronic subsystem is modelled as a continuum heat reservoir, divided into cubic regions on a grid, over which the electronic heat conduction is described by a heat diffusion equation.
We set a grid size similar to that used in previous conventional TTMD simulations in Ni~\cite{Zar17b}, with a grid spacing of \(22~\text{\AA}\).
For the electronic heat capacity, we use the constant value of \(C_\text{e} = \num{1e6}~\mathrm{J m^{-3} K^{-1}}\) \cite{Sam16} and for the electronic thermal conductivity \(\kappa_\text{e} = 88~\mathrm{W m^{-1} K^{-1}}\) \cite{Jin16}.
For the ion-electron coupling in Ni, we use the function presented in \cite{Car19}.

We compare this model to the method commonly used in cascade simulations~\cite{Nor17}, where energetic atoms are subjected to a (scalar) retarding force with a magnitude derived from the electronic stopping power (ESP) given by SRIM~\cite{SRIM}.
The ESP model in SRIM is parametrised by fitting to a large body of experimental data and, thus, is very accurate in the range where data is available.
However, data exist mainly for light ions and become scarce for heavier ions, as well as for low-energy projectiles.
This range of empirical uncertainty coincides with the energy regime relevant to collision cascades.
Electronic stopping theory predicts no lower velocity threshold for metals, and accordingly, SRIM stopping powers are extrapolated down to zero velocity. 
The lower energy region is essentially linear in velocity, with an effective stopping power friction coefficient of \(0.0101~\mathrm{eV ps} / \text{\AA}^2\).
The friction in the traditional MD is solely a function of the velocity, and a lower velocity cut-off is further applied in the MD simulations to avoid unphysical quenching of thermal modes, i.e.,
for particles below the cut-off velocity, the electronic stopping is not applied.
Given that the value of the cut-off is uncertain, we compare predictions using both a \(1\) and a \(10~\mathrm{eV}\) kinetic energy cut-off (denoted by ESP-1eV and ESP-10eV, respectively).

\subsection{Simulation set-up and analysis\label{sec:setup}}

Collision cascades from primary knock-on atoms (PKAs) with  \(E_{\text{kin}} = 50~\mathrm{keV}\) were chosen for this study since this energy lies above the subcascade splitting threshold in Ni~\cite{DeB18}. 
Hence, cascades reach the highest possible energy density so that all relevant physics regimes are captured.
The duration of the thermal spike, and thus the simulated ion beam mixing, is sensitive to the melting point of the interatomic potential~\cite{Nor98d,Nor98}.
We, therefore, use the Foiles EAM potential~\cite{Foi86} for all simulations, which has a melting point within 4\% of the experimental value \cite{Nor98}, and which has been extensively used in cascade simulations~\cite{Nor97f,Nor98,Nor98d,San15}.

Cascades were simulated in cubic atomic systems containing 4.5 million atoms, with periodic boundary conditions applied in all directions.
The cell was initially equilibrated at \(300~\mathrm{K}\), first for \(5~\mathrm{ps}\) with Berendsen scaling thermostat~\cite{Berendsen} to efficiently achieve a temperature equilibrium in the atomic system; then for an additional \(5~\mathrm{ps}\) with the EPH coupling active between the atomic and electronic subsystems, and with the electronic system held at \(300~\mathrm{K}\).
The EPH model recovers the properties of a canonical thermostat under fixed electronic temperature~\cite{Tam18}.
The equilibrated cell was then used as initial configuration for all cascade simulations. 
PKAs were initiated in random directions uniformly distributed over the unit sphere, with positions chosen such that the initial velocity was directed towards the center of the cell.
The borders of the atomic system were monitored to ensure that no energetic atoms crossed the periodic boundary to avoid self-interaction of possibly extended cascades or channeling projectiles.
At least 30 cascades were simulated for any given condition, which provides sufficient statistics to draw quantitative conclusions.

The excess energy introduced by the PKA would, in a real material, diffuse into the surrounding bulk.
In MD simulations carried out in a finite atomic system with periodic boundaries, the heat diffusion into the bulk lattice can be emulated by applying a thermostat on the atoms in the boundary region.
With the EPH model, heat transport occurs mainly via the electronic system, as the electronic heat conduction is orders of magnitude higher than that of the lattice.
By extending the electronic system past the atomic system, the energy introduced by the PKA is allowed to diffuse via the electron thermal conductivity \cite{Duf09}.
The electronic system was extended to a cubic volume with twice the side length of the atomic system, with periodic boundaries also on the electronic system.
No additional thermostat was applied to the atoms other than the coupling from the generalized Langevin dynamics of the EPH model.
The system was large enough that the final temperature only rose by \(19~\mathrm{K}\).
For cascades with the ESP-based models, a Berendsen thermostat was applied to the boundary regions as described in Ref.~\cite{San15}.

We identify residual defects using the Wigner-Seitz (W-S) analysis method implemented in \textsc{Ovito}~\cite{Ovito}.
The simulation cell is divided into Voronoi cells around each lattice site, and empty Voronoi cells are labelled as vacancies, while multiply occupied cells are labelled as interstitials.
Defect clusters were identiﬁed with an automated procedure, where a defect is deﬁned as belonging to a cluster if it lies within a given cut-off radius to another defect in the same cluster.
Clustered self-interstitial atoms (SIAs) lie close together, and the cluster analysis is not sensitive to whether a first nearest neighbor (1NN) or second nearest neighbor (2NN) distance is used.
Vacancy clusters resulting from cascades can be more spread out, and the cluster analysis is thus sensitive to the choice of cut-off.
According to DFT, the 1NN divacancy in Ni is binding, while the 2NN divacancy is repulsive~\cite{Meg10}.
Hence, we use a 1NN cut-off radius for both defect types.

\subsection{Ion beam mixing}

While direct observation of cascade dynamics on the picosecond time scale is not feasible by experimental means, ion beam mixing (IBM) experiments provide a measure of the displacements occurring in the material as a result of the cascade. 
The experimentally measured quantity is based on the widening of a marker layer after ion irradiation~\cite{Pai85}, and directly corresponds to the simulated atomic mixing~\cite{Nor98}. 
The total mixing is the sum of ballistic mixing and thermal spike mixing, the latter being the result of diffusion in the liquid-like core of the cascade in its thermal spike phase. 
Hence, the degree of mixing depends on both the size and the lifetime of the thermal spike and, therefore, provides a useful indication of the accuracy of the simulated cooling rate of the cascade and, thereby, of the validity of the energy loss model for the thermal spike stage.

We calculate the ion beam mixing using the method described in Refs.~\cite{Nor98,San15}, and compare our simulated atomic mixing predictions to the experimental ion beam mixing results of Kim \emph{et al.} \cite{Kim88} and Nordlund \emph{et al.} \cite{Nor98}.
The experiments employed \(600\) and \(650~\mathrm{keV}\) Kr ion beams and were conducted at a temperature of \(6~\mathrm{K}\).

The atomic mixing \(Q\) is given by the equation~\cite{Nor98}
\begin{equation}
  Q = \frac{\int_0^{E_0}{R^2(E)N(E)\mathrm{d}E}}{6nE_{D_n}},
  \label{eq:Q}
\end{equation}
where \(R^2(E)\) is the total atomic displacements squared, \(N(E)\mathrm{d}E\) is the primary recoil energy spectrum from collisions with the incident Kr ions,
\(n\) is the atomic density, 
and \(E_{D_n}\) is the nuclear deposited energy of the ion beam. \(E_0\) is the initial incident Kr ion energy.
The primary recoil spectrum and the nuclear deposited energy were calculated explicitly using the MDRANGE code~\cite{Nor95}, as described in Ref.~\cite{Nor98}.

In order to obtain the displacement function \(R^2(E)\) for the purpose of integration, we performed cascade simulations for a range of PKA energies.
Cascades were simulated at the initial temperature of \(6~\mathrm{K}\), corresponding to the experimental conditions.
We calculated the total atomic displacements \(R^2\) averaged over the cascades for each chosen PKA energy separately.
Displacements due to thermal vibrations were excluded from the calculation by summing only contributions above half the nearest neighbor distance.
We then fit a function to the simulated data points of the formtotal
\begin{equation}
  R^2(E)=a\frac{E^{c+1}}{b^c + E^c},
  \label{eq:R2}
\end{equation}
which has the desired property of becoming linear in energy for higher PKA energies. 
This behaviour is well known~\cite{Nor98} and stems from the subcascade splitting phenomenon~\cite{DeB16}, where cascades from high energy PKAs split into separate smaller regions, and the impact of the cascade becomes a sum of the effects of the smaller cascade regions.
The total simulated ion beam mixing is dominated by the behaviour of \(30-300~\mathrm{keV}\) cascades \cite{Nor98}.
We therefore simulated cascades up to \(200~\mathrm{keV}\) to obtain an extrapolation as accurately as possible to the high energy limit.

\section{Results and discussion}

\subsection{Cascade evolution\label{sec:energyloss}}

\begin{figure}[htp]
\includegraphics[width=1.0\columnwidth]{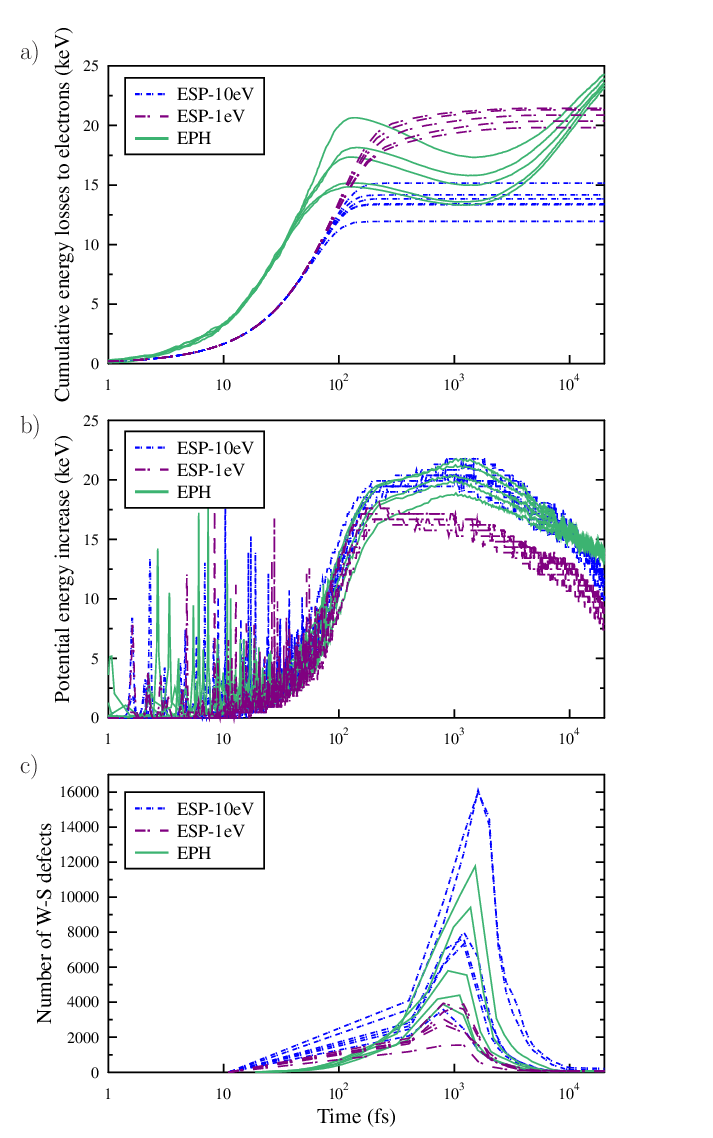}
\caption{
Evolution of the \(50~\mathrm{keV}\) cascades during the first \(10~\mathrm{ps}\).
(a) Cumulative net energy transfer from the atoms to the electronic system.
Note that in two-temperature models such as EPH, electrons can give back energy to atoms;
in contrast, in the ESP-based models, the energy lost to electrons is completely removed from the simulation.
(b) Total potential energy increase in the atomic system.
(c) Disorder in the cascade quantified by a Wigner-Seitz cell analysis.
}\label{fig:evol}
\end{figure}

A dense, energetic collision cascade progresses through three distinct stages of evolution~\cite{Cal10}, which are also clearly visible in the simulations using the EPH model.
The early stage, referred to as the ballistic or destructive~\cite{Cal10} stage, is characterised by a small number of energetic atoms recoiling from strong collisions. The energetic recoils penetrate through the crystal between lattice atoms, increasing the damaged region of the crystal and leading to rapid energy losses to electrons through electronic stopping. The cumulative energy losses are illustrated in Fig.~\ref{fig:evol}a.
The two ESP models predict similar rates of energy losses during the ballistic stage due to the low number of atoms present during that stage with energies in the range between the two cut-off values \(1\) and \(10~\mathrm{eV}\).
However, the energy losses cease earlier with ESP-10eV than with ESP-1eV.
In contrast, the energy losses predicted with the EPH model are more rapid in the early ballistic stage, but also cease earlier than with both ESP models.
Consequently, the cumulative energy losses by the end of the ballistic stage reach a similar level with the EPH model as with the ESP models, albeit with larger statistical variability, with the maximum losses ranging between the levels of the two ESP models.

The evolution of the global potential energy is shown for the three models in Fig.~\ref{fig:evol}b.
During the ballistic stage, the increasing volume of the damaged region results in a rapid increase in the global potential energy.
Despite the differences in energy loss rates noted above, the potential energy shows the same rate of increase with all three models. 
The end of the ballistic stage is marked by a change in the slope of the potential energy curve~\cite{Cal10} and occurs for all models at roughly \(200~\text{femtoseconds}\).
 
After the ballistic stage, the cascade transitions into the thermal spike phase, also referred to as the non-destructive phase~\cite{Cal10}.
At this stage of cascade development, the damaged volume no longer increases (hence the term "non-destructive").
Instead, a disordered and under-dense liquid-like core develops in the cascade region, pressure increases at the outskirts of the cascade, and pressure waves emanate from the core region into the undamaged areas of the material. 

\begin{figure}[h]
\includegraphics[width=1.0\columnwidth]{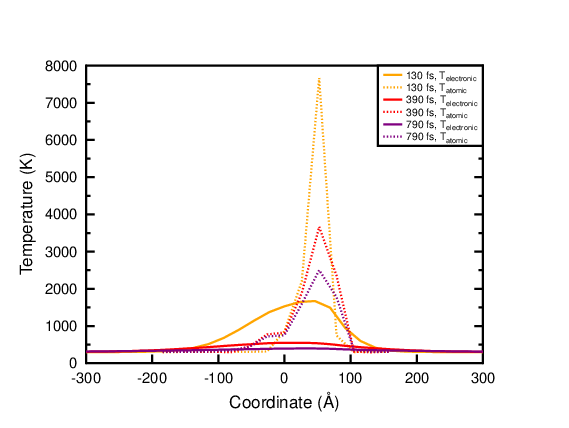} 
\caption{
Temperature profile of the electronic and atomic systems through a \(50~\mathrm{keV}\) cascade, simulated with the EPH model, during the period in time when the energy feedback to the atomic system equals or exceeds the electronic energy losses.}
\label{fig:Tprofile}
\end{figure}

The energy losses to electrons cease completely for the ESP-10eV model at around \(150~\mathrm{fs}\), before the onset of the thermal spike phase. 
In contrast, for the ESP-1eV model, the atomic system continues to lose energy during the thermal spike phase, resulting in an additional loss of \(5-10~\mathrm{keV}\), or in other words, a 50\% higher cumulative energy loss compared to the ESP-10eV model. 
In the EPH model, on the other hand, the electronic system is explicitly modelled, and the coupling is active during the entire simulation, revealing a complex energy coupling between the two subsystems during the thermal spike phase.
We find that the net energy dissipation from the atomic system ceases already after \(100~\mathrm{fs}\).
At this point, the energy transfer to the electronic system levels off and even slightly reverses, with the electronic system feeding energy back into the atomic system.
The balance in energy transfer between the subsystems persists for more than a picosecond, roughly corresponding to the lifetime of the thermal spike. 
The temperature profile through the cascade region is plotted in Fig.~\ref{fig:Tprofile}, and shows that at this stage in the cascade core, the electronic system is already colder than the atomic system.
On the other hand, in the area outside the perimeter of the cascade region, the electrons are hotter than the atoms.
This is the source of the atomic heating seen in the net energy transfer \emph{from} the electrons \emph{to} the atoms, an effect that is not captured in the ESP-models.

Energy transfer between subsystems during the heat spike stage clearly affects the potential energy development in the atomic system (Fig.~\ref{fig:evol}b). 
The global potential energy gradually increases during the thermal spike phase in the ESP-10eV cascade simulations, where energy losses have ceased. 
In contrast, with the ESP-1eV model, where the energy losses continue into the thermal spike phase and reach a much higher level, the potential energy decreases during this phase.
In all EPH-modelled cascades, on the other hand, the potential energy increases during the thermal spike, irrespective of the variation in the total energy losses reached before the onset of the thermal spike stage.
In particular, the results obtained with the EPH model predict a continued potential energy increase during the thermal spike even for cascades where the total energy losses during the ballistic stage reached the same high level as those obtained with the ESP-1eV model, and even when the total energy loss after feed-back remains on a higher level than that predicted with the ESP-10eV model.

We quantify the disorder that develops in the thermal spike region with the W-S analysis, which is also used for identifying the residual defects.
It should be noted that the identified empty and multiply occupied cells during the cascade evolution do not correspond to stable defects, but rather provide a measure of the temporary disturbance of the lattice.
The W-S analysis is not sensitive to thermal displacements, though it does register atoms that are elastically displaced around the thermal spike region as a result of the pressure waves emanating from the core. 
The result of the W-S analysis is shown in Fig.~\ref{fig:evol}c.
The disorder reaches a maximum at the end of the thermal spike phase, around the time when the global potential energy begins to decrease for all models. 
While the time to reach peak disorder depends on the maximum level of disorder that is reached, rather than on the particular energy loss model, a clear model dependence is observed in the height of the maximum. 
The decreasing potential energy found with the ESP-1eV model during the thermal spike phase is reflected in the consistently low peak level of disorder. 
In contrast, with both the ESP-10eV and EPH models, the peak disorder varies strongly, ranging from the maximum level reached by the ESP-1eV cascade of around 4000 W-S defects up to a factor of 3-4 times more.

Most of the damage present at the peak of the disorder recombines during the subsequent cooling and recrystallization stage, leading to the decrease in global potential energy observed after the thermal spike phase with all models.
The higher the peak damage is, the longer it takes for the region to recrystallize.
As the cascade region cools and the temperature continues to even out, the two subsystems in the EPH model continue to equilibrate through the electron-phonon coupling mechanism.
The extended electronic system serves as a heat sink, emulating thermal dissipation into the bulk, which is visible as an increase in the energy transfer to electrons during the cooling phase in  the EPH cascades. 
The corresponding energy dissipation into the bulk, although present in the ESP-model cascades through the boundary thermostat, is not visible in the energy transfer plot in Fig.~\ref{fig:evol}. 
After about \(10~\mathrm{ps}\), the cascade region has recrystallized, and residual defects are essentially stable.
The final remaining defects and defect structures are presented in more detail in Section \ref{sec:defects}.

\subsection{Strength of electron-ion coupling\label{sec:coupling}}

\begin{figure}[h]
\includegraphics[width=1.0\columnwidth]{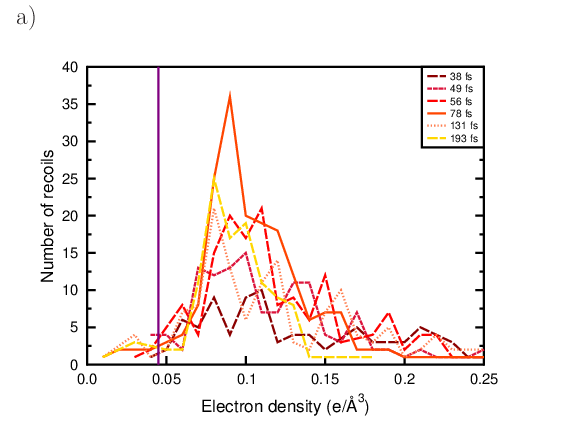}
\includegraphics[width=1.0\columnwidth]{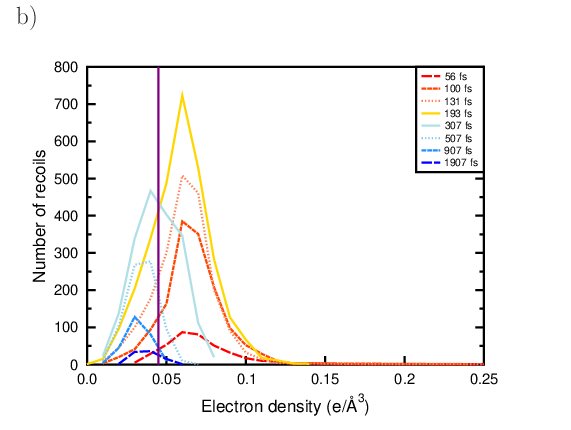}
\caption{
The range of local electronic densities experienced by cascade atoms with energies above \(10~\mathrm{eV}\) (a), and between \(1\) to \(10~\mathrm{eV}\) (b), at different stages of a \(50~\mathrm{keV}\) cascade.
The vertical bar marks the average electronic density experienced by atoms in the undisturbed equilibrium lattice.
Recoil numbers at times during the ballistic stage are plotted in yellow and red hues, while recoils during the thermal spike stage are plotted in blue hues.
}
\label{fig:eldens}
\end{figure}

The rapid electronic energy losses during the ballistic stage of cascade evolution stem from the electronic stopping affecting the energetic cascade atoms. 
In the EPH model, the electronic stopping mechanism is not included as a separate term in the equations of motion, but rather is captured through the strong coupling to the electronic system of atoms in high electron density environments.
The local electron densities explored by cascade atoms in different kinetic energy ranges is plotted in Fig.~\ref{fig:eldens} for various times during the early cascade evolution.
Atoms with energies above \(10~\mathrm{eV}\) (Fig.~\ref{fig:eldens}a), that would experience electronic stopping in both ESP models, consistently see higher local electronic densities than atoms in the equilibrium lattice, in agreement with the expected stronger energy dissipation. 
The numbers of these atoms peak at \(78~\mathrm{fs}\) into the cascade, after which they gradually decrease, and after \(200~\mathrm{fs}\), such high-energy atoms no longer exist.
These atoms are the source of the rapid energy losses predicted by the EPH model in the early ballistic stage.

Figure~\ref{fig:eldens}b shows the local electron densities experienced by atoms with kinetic energies between \(1\) and \(10~\mathrm{eV}\).
Atoms in this energy range exist for up to \(2~\mathrm{ps}\) after the initial recoil event, and are subject to the electronic stopping friction force in the ESP-1eV model but not with the ESP-10eV model.
In the EPH model, the local electronic density visited by these atoms, and hence the coupling strength that they experience, changes as the cascade transitions from the ballistic stage to the thermal spike stage.
In the ballistic stage, these atoms experience environments with a higher electronic density and stronger coupling than bulk atoms, contributing to the electronic stopping energy losses.
However, as the cascade transitions into the thermal spike stage, we observe a shift to environments with densities that lie below that of atoms in the crystalline lattice sites.
The low electron densities exist in the center of the hot, under-dense cascade core, thus atoms there are only weakly coupled to the electronic system. 

\begin{figure}[h]
\includegraphics[width=1.0\columnwidth]{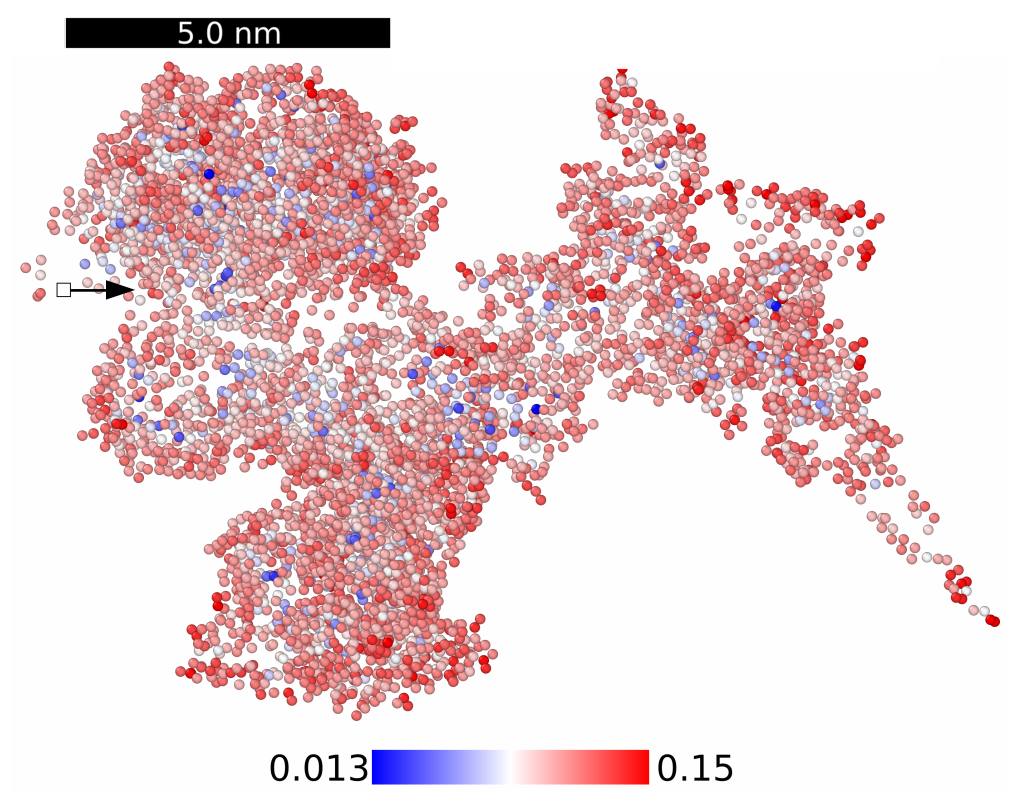}
\caption{
Atoms with \(E_\text{kin} > 0.5~\mathrm{eV}\) (\(v > 0.128~\text{\AA/fs}\)) at \(112~\mathrm{fs}\) into a \(50~\mathrm{keV}\) cascade, coloured on logarithmic scale according to their local electronic density, adjusted so that the average electronic density for atoms in equilibrium (\(0.045~e/\text{\AA}^3\)) lies in the white region.
Red indicates a higher electron density and, hence, a stronger coupling (strong friction and feedback), while blue indicates a lower electron density than that experienced by equilibrium atoms. 
The initial position and direction of the PKA are indicated by the box and black arrow.
}
\label{fig:Frame}
\end{figure}

Figure~\ref{fig:Frame} illustrates the range of the local electron density experienced by atoms in different regions of a cascade in a snapshot taken \(100~\text{femtoseconds}\) into the cascade development. 
Only atoms with kinetic energy above \(0.5~\mathrm{eV}\) are shown, and the color scale has been adjusted so that white indicates the level of the local electron density of atoms in equilibrium lattice conditions (\(0.045~e/\text{\AA}^3\)).
The cascade was initiated by a PKA near the top left-hand corner of the image, directed towards the right.
The highest electron densities are found for atoms around the outskirts of the cascade, where ballistic recoils are still penetrating regions of undisturbed crystal. 
The local environment of these recoils is consistent with the picture of a projectile travelling through a pristine lattice, which underlies the LSS-theory~\cite{Lin63} of electronic stopping. 
This is captured by the strong coupling in the EPH model.
In contrast, atoms behind the cascade front, in the core of the cascade, are not traveling through intact crystal, but rather collectively form a disordered, under-dense region. 
This leads to a low local electronic density (blue color in Fig.~\ref{fig:Frame}), and hence a weak coupling to the electronic system according to the electron density dependent dissipation model. 

\begin{figure}
\includegraphics[width=1.0\columnwidth]{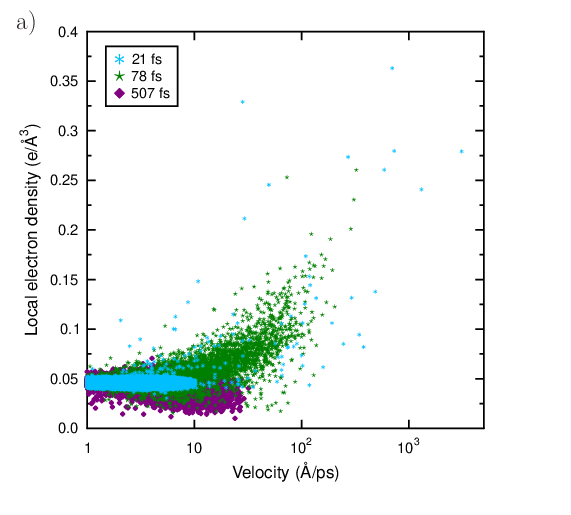}
\includegraphics[width=1.0\columnwidth]{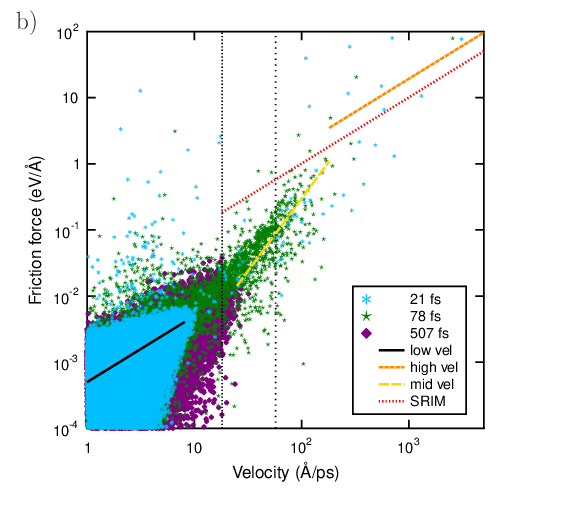}
\caption{
(a) The local electron density vs. the instantaneous kinetic energy of all atoms at \(21~\mathrm{fs}\), \(78~\mathrm{fs}\), and \(507~\mathrm{fs}\) into the cascade (see the main text for a detailed description).
(b) Retarding component of the friction force \(-\mathbf{\sigma} \cdot \mathbf{\hat v} \) for all atoms at \(21~\mathrm{fs}\), \(78~\mathrm{fs}\), and \(507~\mathrm{fs}\) simulated with the EPH model. Lines show fits to data in different velocity regimes, indicated by the extent of the line (see text for details).
The equivalent of the ESP model (based on SRIM stopping power) is also plotted, with the \(1~\mathrm{eV}\) and \(10~\mathrm{eV}\) cutoff values of the ESP models indicated with vertical dotted lines.
}
\label{fig:coupling}
\end{figure}

Figure~\ref{fig:coupling}a shows the local electron density for all atoms in the simulation cell at three different times during a cascade simulation.
For visualization purposes, we plot only the range of velocities between \(1-5000~\text{\AA/ps}\), leaving out the lower range of the atoms near equilibrium conditions.
At \(21~\mathrm{fs}\) into the cascade, the range of electron densities for atoms in the undisturbed lattice is clearly visible as the compact blue region centered on \(0.045~\mathrm{e} / \AA^3\).
The outliers represent the early stage energetic recoils, visiting regions of undisturbed crystal and thus penetrating between atoms in equilibrium lattice sites.
The recoil energy is at times transformed into potential energy; hence, a number of atoms appear with low velocity yet high electron density at this stage.
At \(78~\mathrm{fs}\), the ballistic phase has progressed further, and many more atoms have higher energies and are exploring regions of higher electron density, which is seen by the upward arch of the green body of markers. 
At \(0.5~\mathrm{ps}\), well into the thermal spike phase, atoms in the cascade core are still hot, with kinetic energies above the thermal level. 
However, a large fraction of the atoms, including those with high kinetic energies, experience a lower electron density than atoms in the equilibrium lattice (seen by the downward dip of the purple body of markers in Figure~\ref{fig:coupling}a), effectively limiting the energy flow out of the thermal spike.

Figure ~\ref{fig:coupling}b shows the non-adiabatic force on the individual atoms resulting from the coupling in the EPH model. 
Linear fits to the low-velocity region (\(E_\text{kin} < 0.2~\mathrm{eV})\), corresponding to the bulk of the atoms in the undisturbed lattice, and to the high energy region (\(E_\text{kin} > 100~\mathrm{eV}\), give effective friction coefficients (slopes) of \(0.0005~\mathrm{eV ps / Å^2}\) and \(0.0193~\mathrm{eV ps / Å^2}\) respectively. 
This can be compared to the single-slope \(0.0101~\mathrm{eV ps/ Å^2} \) of the SRIM model (yellow line in the figure).
Normalizing by the mass of the nickel atoms, this gives two separate characteristic times emerging from the EPH model, \(12~\mathrm{ps}\) for the electron-phonon coupling regime and \(0.3~\mathrm{ps}\) for the ballistic regime.
Fitting to a general power law \(\propto v^m\) in these ranges yields exponents of \(0.987 \pm 0.006\) and \(1.169 \pm 0.122\), respectively, so we conclude that the underlying data is, in effect, strongly linear despite the large spread.
In the intermediate energy range, however, a linear fit does not capture the crossover between the two regimes;
we find a fit to a power law \(\propto v^m\) with \(m = 2.191 \pm 0.355\) in this region.
This exponent is an intrinsic property of the non-equilibrium atom configurations and the model's geometry.
The degree of universality of this proposed power law, when applied to other systems and conditions, remains to be evaluated.

\begin{figure*}[htp]
\includegraphics[width=2.\columnwidth]{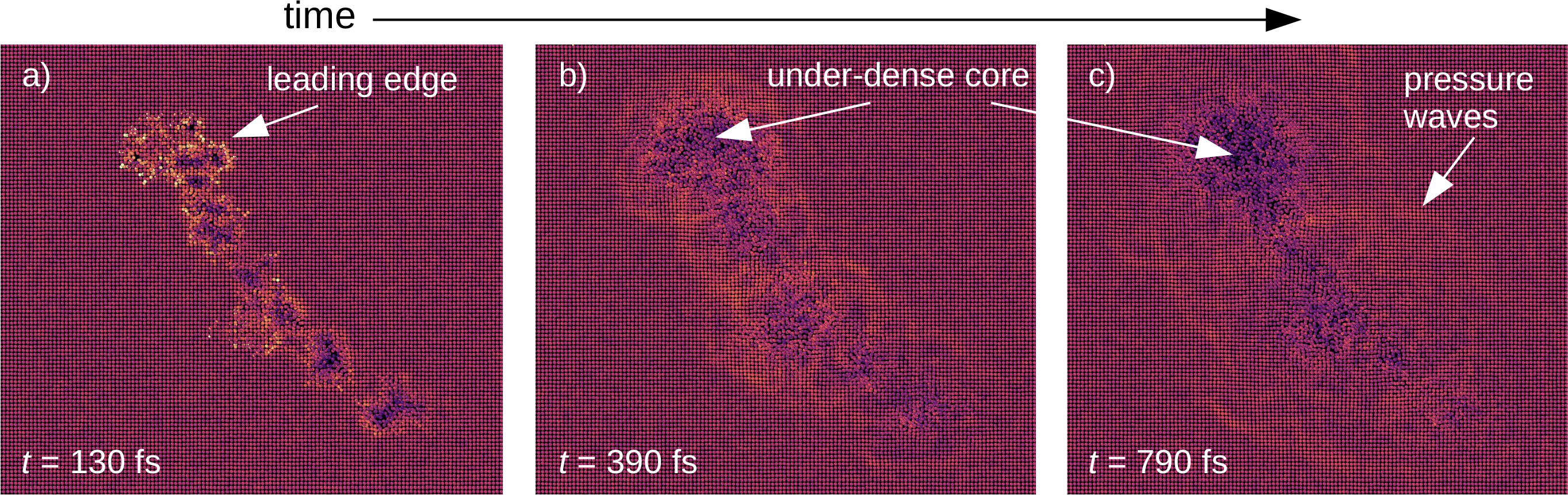}
\caption{
The individual coupling strength \(\alpha ^2 (\rho)\) for atoms in a two unit cell thick slice through a \(50~\mathrm{keV}\) cascade.
The color scale represents the strength of the coupling, which is correlated with local compression.
Values of \(\alpha ^2 (\rho)\) range from 0.001 \(\mathrm{eV ps} / \mathrm{\AA}^2\) (black) to 0.01 \(\mathrm{eV ps} / \mathrm{\AA}^2\) (yellow) on a logarithmic scale.
}
\label{fig:slice}
\end{figure*}

The spread in the values of local electron densities of the lattice atoms is much wider during the thermal spike phase than in the early ballistic stage (cf. the dense bodies of blue and purple markers in Figure ~\ref{fig:coupling}a).
The larger variation in the later stage is due to the strong pressure and density gradients that develop in and around the disordered cascade core, leading to a range of coupling parameter values throughout the material in proximity to the cascade.
The full atomic system in a slice of material through the cascade core is illustrated in Fig.~\ref{fig:slice}.
The coupling strength increases with increasing density in the region around the cascade core, as well as in the pressure waves that emanate from the cascade core.
In frame (a), the cascade is nearing the end of the ballistic stage.
The recoil was initiated in the lower right-hand corner, and small pockets of local disorder are beginning to form in this region. 
The surrounding lattice is undisturbed.
In the top region, the cascade is still expanding, and one can see the ballistic atoms with strong coupling (yellow) in the leading edge.
At \(0.4~\mathrm{ps}\), in frame (b), a thermal spike region has begun to form at the top of the image. 
In frame (c), the disordered core of the thermal spike is well developed, and the coupling in the core is very weak. 
Pressure waves emanating from the cascade region are clearly visible as bands of more strongly coupled atoms due to the correlation of the coupling parameter \(\alpha^2(\rho)\) with local compression.

At the time of the snapshots in (b) and (c), the electronic temperature in the cascade core has already decreased to around \(400-500~\mathrm{K}\) (see Fig. \ref{fig:Tprofile}).
However, through the electronic heat conductivity, the electronic temperature in the region surrounding the thermal spike has risen above the lattice temperature.
The stronger coupling in the high-pressure regions around the core thus leads to efficient heating of the atomic system by the electronic system in the EPH model, resulting in the dip in the cumulative energy transfer seen during the thermal spike phase (Fig.~\ref{fig:evol}).

\subsection{Damage formation\label{sec:defects}}

\begin{figure}[htp]
\includegraphics[width=1.0\columnwidth]{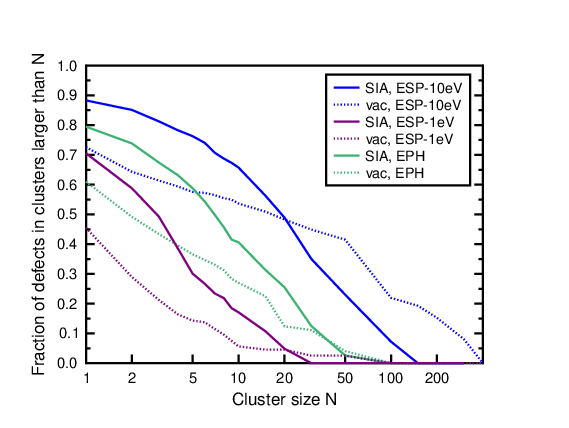}
\caption{
The fraction of surviving SIAs and vacancies found in clusters larger than size \(N\) as a function of \(N\).
Results are averaged over 30 different cascades for each case.
}
\label{defclust}
\end{figure}

The average numbers of surviving defects and the largest cluster sizes for cascades simulated with the different dissipation models are given in Table~\ref{table:def}. 
A high level of peak disorder (Fig.~\ref{fig:evol}c) is found to correlate with a larger number of residual defects and with the formation of large defect clusters.
Characterizing cluster sizes in the primary damage is of importance for understanding the longer scale defect evolution since defect clusters migrate and interact differently from point defects, resulting in qualitative and quantitative differences in damage accumulation~\cite{Mas14}.
A larger fraction of the surviving defects of both SIA and vacancy type are predicted to belong to clusters in the ESP-10eV cascades, while the ESP-1eV method predicts a larger fraction of the final damage to be in the form of point defects, with the predictions of the EPH model lying in between, as shown in Figure~\ref{defclust}.
For the ESP-1eV cascades, which never reach a high level of disorder, large clusters were not seen at all. 
Since there is no \emph{a priori} way of determining a reasonable cut-off velocity for the electronic stopping in the ESP models, effectively, the EPH model reduces the theoretical uncertainty in Frenkel pair numbers (from \(60-107\) to \(73 \pm 4\)) stemming from arbitrary choices of the cut-off, and refines the statistics of cluster sizes.

\begin{table}
\caption{
Statistics of defects in \(50~\mathrm{keV}\) cascades simulated at \(300~\mathrm{K}\) with different dissipation models.
The average numbers of Frenkel pairs (FP), single vacancies (V), and single self-interstitial atoms (SIA) are given, averaged over 30 individual cascades for each case. 
The size of the single largest vacancy cluster (\(Cl_{\text{V}}^{max}\)) and SIA cluster (\(Cl_{\text{SIA}}^\text{max}\)) found in any cascade is also given.
}
\centering
\begin{tabular}{l c c c c c}
\hline
\hline
Model     & FP              & single V     & single SIA     & \(Cl_{\mathrm{V}}^{max}\) & \(Cl_{\mathrm{SIA}}^{max}\) \\
\hline
EPH       & \( 73  \pm  4\) & \(28 \pm 1\) & \(15 \pm 1\)   &  89                       &  56                         \\
ESP-10eV  & \(107  \pm 13\) & \(29 \pm 1\) & \(12 \pm 1\)   & 307                       & 136                         \\
ESP-1eV   & \( 60  \pm  2\) & \(33 \pm 1\) & \(18 \pm 1\)   &  54                       &  30                         \\
\hline
\hline
\label{table:def}
\end{tabular}
\end{table}

The trend between the EPH model and the ESP-10eV model in our work is similar to that found in Ref.~\cite{Zar16} between the conventional TTMD model and the model using electronic stopping with a cut-off value equal to twice the cohesive energy (\(2\times 8.88~\mathrm{eV}\)) of Ni, although the interatomic potential we use for the pure Ni system predicts fewer defects on average than the potential used in Ref.~\cite{Zar16}.
We note that, in general, a potential developed for an alloy is not necessarily optimal for the single-element system; hence, differences can be expected.
There is also a similar trend in the fraction of clustered defects, as well as the sizes of the largest SIA and vacancy clusters, which decreases in the EPH and conventional TTMD models compared to the models without an electronic system that use a relatively high electronic stopping threshold.

\subsection{Experimental validation of mid-energy range coupling: Ion beam mixing}

Whereas the predictions of the EPH model can be directly compared to theoretical models in the electronic stopping (high energy) and electron-phonon coupling (low energy) regimes~\cite{Car15}, the thermal spike dynamics are not well described by the underlying assumptions of either the theory of electronic stopping or that of electron-phonon coupling.
Nevertheless, the development of the thermal spike is sensitive to the energy loss model applied in simulations. 
We therefore validate the EPH model in this mid-energy regime by evaluating the effect of the thermal spike on the atomic system. 
The thermal spike results in atomic mixing, which, in terms of the average atomic displacements \(R^2\), is directly comparable to ion beam mixing experiments. 
Although experimental ion beam mixing values are subject to a high degree of uncertainty, compounded by effects of chemical driving forces~\cite{Pai85} not included in these simulations, they represent one of the few quantitative predictions of the direct impact of collision cascades on the atomic scale.

\begin{figure}[h]
\includegraphics[width=1.0\columnwidth]{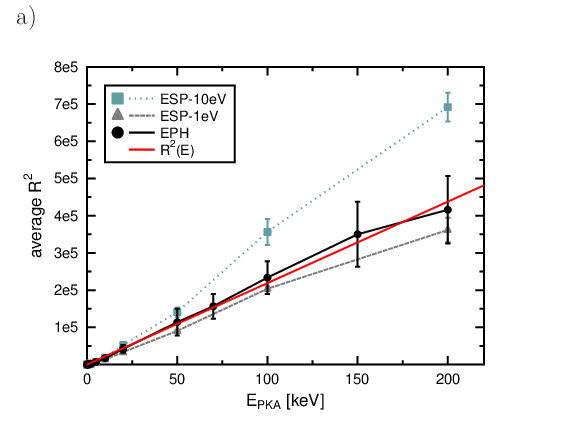}
\includegraphics[width=1.0\columnwidth]{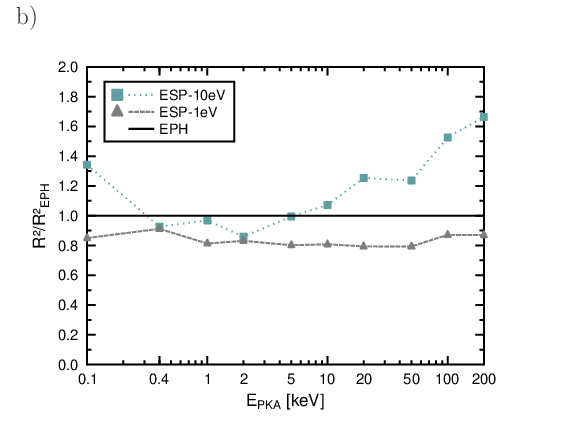}
\caption{
(a) Average displacement \(R^2\) in cascades simulated with the three different methods of treating the electron-ion coupling (see text for details). 
The function (\ref{eq:R2}) for \(R^2(E)\) fitted to the EPH model data is also shown. 
(b) The ratio of \(R^2\) simulated with the two ESP-based models to the values obtained with the EPH model as a function of PKA energy (logarithmic scale).
}
\label{r2lin}
\end{figure}

The simulated values we obtain for \(R^2\), together with the fitted function (\ref{eq:R2}), are shown in Fig.~\ref{r2lin}a.
For comparison, \(R^2\) values obtained in Ref.~\cite{San15} from cascade simulations with the ESP models are included in the plot as well.
We obtain the primary recoil spectrum \(N(E)\mathrm{d}E\) for the Kr ions with \(600~\mathrm{keV}\) and \(650~\mathrm{keV}\) energy in a depth range of \(200-600~\mathrm{nm}\) (centered around the depth of the marker layer in the experimental sample) using the molecular dynamics-based ion range code MDRANGE \cite{Nor95}.
The results of integrating function (\ref{eq:Q}) are given in Table \ref{table:mixing}.
Considering the fact that the EPH model contains no adjustable free parameters, in contrast to the ESP models, our results are in good agreement with the experimental values.

\begin{table}
  \caption{Simulated values for ion beam mixing (IBM), compared to experiment, and to predictions with other electronic stopping methods.}
  \centering
  \begin{tabular}{l c c}
    \hline
    \hline
    Model                                    & \(600~\mathrm{keV}\) Kr & \(650~\mathrm{keV}\) Kr \\
    \hline
    EPH \((\text{\AA}^5/\mathrm{eV})\)       & \(3.8 \pm 0.1\)         & \(4.0 \pm 0.1\)         \\
    ESP-10eV \((\text{\AA}^5/\mathrm{eV})\)  & \(4.7 \pm 0.1\)         & \(5.1 \pm 0.1\)         \\
    ESP-1eV \((\text{\AA}^5/\mathrm{eV})\)   & \(2.9 \pm 0.1\)         & \(3.1 \pm 0.2\)         \\
    \hline
    expt. \((\text{\AA}^5/\mathrm{eV})\)     & \(4.8 \pm 0.5\)         & \(5.0 \pm 0.7\)         \\
    \hline\hline
    \label{table:mixing}
  \end{tabular}
\end{table}

Figure~\ref{r2lin}b shows the ratio of the \(R^2\) values obtained with the ESP models to that obtained with the EPH model for different PKA energies.
The ESP-1eV model consistently gives lower atomic mixing for all PKA energies by a constant factor of 0.8 compared to the EPH model prediction.
This is true despite the initially more rapid energy losses, as observed for the \(50~\mathrm{keV}\) cascades, and it is consistent with the observation that the peak damage level during the thermal spike in \(50~\mathrm{keV}\) EPH cascades lies above that of the ESP-1eV cascades.
The relation between the EPH and the ESP-10eV cascades is more complex.
The agreement is fairly close for lower energy cascades, although even the ESP-10eV model predicts less atomic mixing than the EPH model for the low-energy PKAs.
In contrast, for increasing PKA energies above \(10~\mathrm{keV}\), the ESP-10eV model increasingly over-predicts the atomic displacements compared to the EPH model. 
The reason for the observed differences in different PKA energy ranges lies in the thermal spike phase of the cascade development.
For PKA energies below \(10~\mathrm{keV}\), a sustained disordered thermal spike does not develop, and the contribution to the atomic mixing stems mainly from ballistic mixing. 
Interestingly, this indicates that the ballistic mixing is slightly higher with the EPH model despite the more rapid energy losses observed during the ballistic stage.
For higher PKA energies, a thermal spike develops and lasts up to several picoseconds.
Hence, the treatment of the energy losses affecting the thermal spike in the different models has a strong effect on the predictions of atomic mixing in the higher energy range.

\section{Discussion}

The TDDFT-calculated electronic stopping, as parameterized in the EPH model, predicts a higher energy loss rate than the ESP-based models during the early ballistic stage in \(50~\mathrm{keV}\) cascades.
We note that this does not necessarily imply that the EPH model is incorrect, since experimental data, to which the SRIM stopping power is fitted, are typically obtained with light ions, or at much higher projectile energies than the recoil energies occurring in irradiation-induced collision cascades.
Furthermore, the frequent strong collisions and momentum transfer that take place during the ballistic stage of the cascade do not occur in experiments designed to measure electronic stopping, as they purposely minimize the nuclear stopping.
Hence, although the electronic stopping given by SRIM is considered valid for random trajectories of energetic projectiles, it is not obvious which model provides the more accurate prediction for electronic stopping energy losses during the ballistic stage of a collision cascade.

In the ESP-based energy loss models (incorporated in simulations without an electronic system), it is implicitly assumed that the energy that feeds into the electrons through the electronic stopping diffuses rapidly enough to be essentially removed instantaneously, and thus, it is expected to no longer affect the evolution of the cascade.
We tested this hypothesis with the explicit electronic conduction in the EPH model, where electrons are not just a sink of energy but a thermal reservoir.
In fact, we find that the energy of the electronic system is retained locally long enough to affect the global energy transfer between subsystems.
The feedback of energy occurring during the thermal spike phase of the cascade has a decisive impact on the potential energy development, also affecting the final defect formation.

Nevertheless, the temperature profile through the center of the simulation cell (Fig.~\ref{fig:Tprofile}) shows that the electronic system in the core region of the cascade cools rapidly enough that extreme temperatures exist for a negligibly short time compared to the evolution of the cascade.
Hence, we expect that the temperature dependence of the electronic heat capacity, which becomes significant at electronic temperatures on the order of thousands of degrees~\cite{Lin08}, would not significantly affect the damage prediction.

\section{Conclusion}

The EPH model treatment of the electron-ion coupling in two-temperature molecular dynamics provides a new tool that is exceptionally well suited for simulating dynamic, far-from-equilibrium processes such as radiation-induced collision cascades.
With the EPH model, the effect of local electronic energy dissipation is captured based on predictions from TDDFT, which are superior in detail compared to homogeneous electron gas models and have been shown to provide good agreement with experimental measurements when combined with MD; both in the high-energy electronic stopping \cite{Lee20} and low-energy electron-phonon coupling \cite{Tam18} limits.
Here, we validate the dissipation model for non-equilibrium dynamics during radiation-induced collision cascades where, in particular, neither the electron-phonon coupling nor electronic stopping theories directly apply.
We employ an existing first principles parametrization for nickel \cite{Tam19}, and compare predictions of collision cascade simulations to experimental ion beam mixing measurements, which are sensitive to the energy landscape of the disordered cascade core and to the rate of energy losses and cooling.

During the cascade dynamics, atoms experience energies ranging over several orders of magnitude, and the material exhibits strong pressure and temperature gradients, as well as under-dense, disordered, and ordered phases of the material.
The electronic energy losses from the atomic system in cascade simulations with the EPH model clearly show the emergence of two different regimes of energy losses under these complex conditions, corresponding, respectively, to the early stage of cascade evolution where electronic stopping is applicable, and to the later stages where the electron-phonon coupling is active.
The two regimes are characterised by different rates of energy losses (corresponding to the characteristic times (\(0.3~\mathrm{ps}\) and \(12~\mathrm{ps}\) respectively) as expected from the different effective coupling strengths of the theories of electronic stopping and electron-phonon coupling.
The different energy loss regimes are separated in time by an intermediate steady state, where the energy transfer between the atomic and electronic subsystems balances out.
The steady state occurs during the thermal spike phase of the cascade, where neither the theoretical model of electronic stopping nor equilibrium electron-phonon coupling fully applies.
These two regimes are also separated in velocity by a non-linear intermediate regime.
We propose that there exists a regime where dissipation effectively corresponds to individual forces that are quadratic on the velocity (exponent near \(2\) in a power law fitting).
Predictions with the EPH model of ion beam mixing, which is sensitive to the development of the thermal spike, are found to be in good agreement with experimental values.

Both the total defect numbers and the frequency of large defect clusters predicted by the EPH model lie between the predictions of the two ESP models.
This further strengthens confidence in the accuracy of the EPH model, since the traditional ESP models are found in many cases to give good agreement with experiments.
However, the absence of an \emph{ad hoc} threshold on the stopping power and an effectively parameter-free treatment of the coupling (independent also of the SRIM model predictions) means that simulations with the EPH model are genuinely predictive.
The EPH method thus provides a significant improvement upon models currently in use by removing the arbitrariness of the cut-off values for coupling constants also employed in conventional two-temperature models, and thus reducing the uncertainty in predictions of the primary damage from collision cascades.

\section*{ACKNOWLEDGMENTS}

The work of AES was partly supported by the Academy of Finland through grant number 311472. 
GPK acknowledges support from the Aalto Science Institute of the Aalto University School of Science.
AAC work was supported by the Center for Non-Perturbative Studies of Functional Materials Under Non-Equilibrium Conditions (NPNEQ) 
funded by the Computational Materials Sciences Program of the US Department of Energy, Office of Science, Basic Energy Sciences, Materials Sciences and Engineering Division.
AT and AAC work was performed under the auspices of the US Department of Energy by Lawrence Livermore National Laboratory under Contract DE-AC52-07NA27344.
Computational resources for this work came from the Aalto University School of Science ``Science-IT'' project, and from CSC – IT Center for Science, Finland.

\bibliography{andreasbib}

\end{document}